# In-poor IGZO: superior resilience to hydrogen in forming gas anneal and PBTI

Anastasiia Kruv*, Michiel van Setten, Adrian Chasin, Daisuke Matsubayashi, Hendrik Dekkers, Alexandru Pavel, Yiqun Wan, Kruti Trivedi, Nouredine Rassoul, Jie Li, Yuchao Jiang, Subhali Subhechha, Geoffrey Pourtois, Attilio Belmonte, and Gouri Sankar Kar

imec, Leuven, Belgium, *email: anastasiia.kruv@imec.be

**KEYWORDS**: DRAM, forming gas anneal, hydrogen, IGZO, oxide semiconductors, reliability

**ABSTRACT** Integrating In-Ga-Zn-oxide (IGZO) channel transistors in silicon-based ecosystems requires the resilience of the channel material to hydrogen. Standard, In-rich IGZO (In = 40 metal at. %), suffers from degradation under forming gas anneal (FGA) and hydrogen driven positive bias temperature instability (PBTI). In this paper we demonstrate scaled, top-gated transistors with an atomic layer deposition (ALD) deposited In-poor (In ≤ 17 metal at. %) IGZO channel that show superior resilience to hydrogen compared to those with an In-rich IGZO channel. These devices, fabricated with a 300-mm semiconductor fabrication plant (FAB) process, with dimensions down to $W_{CH} \times L_{TG} = 80 \times 40$ nm$^2$, show excellent stability in 2-hour, 420ºC forming gas anneal ($0.06 \leq |\Delta V_{TH}| \leq 0.33$V) and improved resilience to H in PBTI at 125ºC (down to no detectable H-induced $V_{TH}$ shift) compared to In-rich devices. We demonstrate that the mechanism of device degradation by H in the FGA is different from that of the H-induced $V_{TH}$ instability in PBTI. We argue the first to be the oxygen scavenging by H and the second H release from a gate-dielectric into the channel. We also show that resilience to H in one process does not automatically translate to resilience to H in the other one. This significant improvement in IGZO resilience to H enables the use of FGA treatments during fabrication needed for silicon technology compatibility, as well as further scaling and 3D integration, bringing IGZO-based technologies closer to mass production.

## I. INTRODUCTION

In-Ga-Zn-oxide (IGZO) transistors have garnered significant attention across a wide range of applications, including display[1,2], back end of line (BEOL)[3,4], radio frequency (RF) circuits[5], neuromorphic computing[6,7], flexible electronics[8,9], flash memory[10], and dynamic random access memory (DRAM)[11]. This interest is driven by the attractive set of IGZO properties, such as its extremely low off-current (down to $10^{-22}$A/μm)[12], reasonable mobility (> $10cm^2/(V*s)$)[2], low-temperature deposition (down to 25ºC)[13], and industrial manufacturability. However, the widespread industrial adoption of IGZO is significantly limited by its extreme sensitivity to hydrogen (H)[14,15]. Exposure to H and its incorporation in the deposited films occur during fabrication, as H is present in the ambient due to the precursors used[16]. Moreover, IGZO-based device integration might adopt forming gas anneal (FGA), which is widely used in Si-cointegration to improve dielectric quality and usually performed for 30min at 420ºC in diluted H environment[14,17]. This step can further exacerbate the issue of the IGZO sensitivity to H.

The interaction between IGZO and H has a profound impact on both device performance and reliability. From a performance perspective, H acts as an n-type dopant in IGZO[8,14,15,18], which boosts the device's on-current ($I_{ON}$) but causes an undesirable negative shift in the threshold voltage ($V_{TH}$)[19]. To bring $V_{TH}$ to application-specific values (such as 0V for DRAM), additional annealing in an oxygen environment is required[11,20]. This additional step not only increases fabrication time and costs but also necessitates the integration of an oxygen tunnel (OT)[21], which is incompatible with the future 3D integration. Furthermore, IGZO's exposure to H at elevated temperatures risks channel dissociation[22], which constrains the overall processing thermal budget. On the reliability front, H release from the gate dielectric into the channel results in negative $V_{TH}$ shifts, as observed in positive bias temperature instability (PBTI) tests, posing a significant challenge to device reliability[15,23].

Previous studies explored several ways of enhancing the resilience of IGZO to H. Those include the use of H blocking layers (e.g., metal, $Al_2O_3$, $SiO_2$, $Si_3N_4$)[8,14,18,16], fluorine (F) doping[8,14,18], and adoption of crystalline IGZO[22]. However, these approaches require extra steps and have several other limitations. For example, deposition of the encapsulation layers themselves can induce H or device shorts[8], F doping can cause dielectrics corrosion[8], and IGZO crystallization at 550ºC increases thermal budget limiting the BEOL compatibility[22].

In this work, we show that IGZO's intrinsic resilience to H can be strongly improved by using In-poor (In ≤ 17 metal at. %) compositions. This composition range is rather unexplored, as most of the studies focus on In-rich films to achieve higher mobility[2,24,25,26,27,28,29,30,31,32,33,34,35]. However, we experimentally show in this work that scaled top-gated (TG) transistors with In-poor IGZO channels can offer reasonable mobility, as well as provide superior resilience to H in FGA and PBTI compared to the In-rich counterpart. Moreover, we demonstrate that resilience to H in those two processes is guided by different mechanisms, and resilience in one does not necessarily translates into resilience in another.

## II. EXPERIMENTAL SECTION

### A. Sample fabrication

The experiments were performed on scaled (down to $W_{CH}$ x $L_{TG}$ = 80 x 40 $nm^2$) top gate (TG) field-effect transistors (FET), fabricated using the gate-last process in a 300 mm FAB[36], see Fig. 1. The FET consisted of a 5 nm ALD $Al_2O_3$ gate dielectric and 7 nm ALD IGZO channel featuring six different compositions (see Tab. 1). The compositions are verified using X-ray Fluorescence (XRF). The compositions are given as the ratio of In, Ga, and Zn to the total number of metal atoms, e.g., without accounting for oxygen. The device source and drain contacts were formed using 10 nm CAAC (c-axis aligned) IGZO raised contacts[37], a 10 nm TiN liner, and a W fill followed by chemical-mechanical polishing (CMP). The insertion of SiN and $SiO_2$ layers beneath the channel forms an oxygen tunnel (OT), which promotes the diffusion of oxygen towards the channel region while limiting its migration toward the contact areas. The $O_2$ anneal is employed to induce $V_{TH}$ shifts in as-fabricated devices, as it allows to passivate oxygen vacancies and reduce the channel doping[21]. A 1 h, 250ºC $O_2$ anneal was used in case of In-rich $In_{0.40}Ga_{0.35}Zn_{0.25}$, but not for the In-poor samples, which already had positive $V_{TH}$ directly after the fabrication. Further details on the sample fabrication can be found in the Supporting Information.

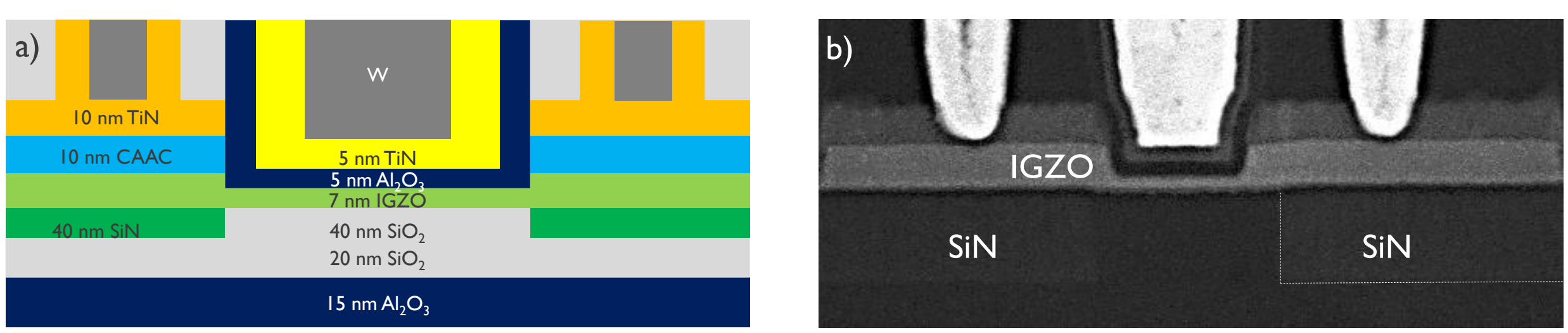

Fig. 1: (a) A schematic illustration (not to scale) and (b) TEM image of the studied devices.

Tab. 1: IGZO compositions of the tested samples and corresponding $O_2$ anneals. In, Ga, Zn concentrations are given relative to the total number of metal atoms, excluding the O content.

| Composition | $In_{0.05}Ga_{0.45}Zn_{0.50}$ | $In_{0.05}Ga_{0.35}Zn_{0.60}$ | $In_{0.10}Ga_{0.45}Zn_{0.45}$ | $In_{0.10}Ga_{0.35}Zn_{0.55}$ | $In_{0.17}Ga_{0.45}Zn_{0.38}$ | $In_{0.40}Ga_{0.35}Zn_{0.25}$ |
|---|---|---|---|---|---|---|
| $O_2$ anneal | No | | | | | 1h 250 ºC $O_2$ |

B. Electrical performance characterization

The electrical performance of the devices was assessed by measuring their transfer characteristics ($I_D$-$V_G$). The following biases were applied to the source ($V_S$), drain ($V_D$), bulk ($V_B$), and top gate ($V_G$) electrodes: $V_S = 0$ V, $V_D = 0.05$ V or 1 V, $V_B = 0$ V, $V_{TG} = [-2; 3]$ V. Up to 20 devices per channel composition per dimension were tested. The reference dimension was chosen to be $W_{CH}$ x $L_{TG}$ = 80 x 40 $nm^2$ to focus on scaled devices that are the most relevant for future industrialization. The following parameters were extracted from the transfer characteristic: field-effect mobility (μ), $V_{TH}$, $I_{ON}$, subthreshold swing (SS), and contact resistance ($R_C$). Mobility was calculated as $\mu = \frac{gm_{MAX} * L_{TG}}{W_{CH} * C_{OX} * V_D}$ for $W_{CH}$ x $L_{TG}$ = 1 x 1 $\mu m^2$, where $gm_{MAX}$ is the maximum transconductance, $V_D = 0.05$ V, and $C_{OX}$ is gate oxide capacitance per area (extracted experimentally). $V_{TH}$ was calculated using a constant current method at $I_D = 10^{-10} * \frac{W_{CH}}{L_{CH}}$ A, $I_{ON}$ was extracted as $I_D$ at $V_G = V_{TH} + 1V$, and SS was defined as the minimum value of the subthreshold swing (calculated as $\frac{dV_G}{dI_D}$ for $I_D > 5 * 10^{-13}$A). $V_{TH}$, $I_{ON}$, and SS were extracted at $V_D = 1$ V. $R_C$ was extracted via the transmission line method[38] at $V_G = V_{TH} + 1$ V, $V_D = 1$ V for devices with source-drain distance $L_{SD}$ = 70-200 nm, $W_{CH}$ = 80 nm. In this method, the total output resistance of a transistor ($R_T = \frac{V_D}{I_D}$) was plotted as a function of $L_{SD}$ and this dependency was fitted with a linear line, intercept of which resulted in 2 $R_C$.

C. Concurrent thermal stressing

The device resilience to FGA was evaluated in concurrent thermal stress tests, where all samples were annealed simultaneously to exclude the impact of the annealing process variation. The anneals were performed for 1 h + 1 h at 420ºC and 1 h + 1 h at 500ºC in a 10% $H_2$ + 90% $N_2$ environment. A fresh set of samples was used for each annealing temperature. The transfer curves were measured before and after annealing for devices with $W_{CH}$ x $L_{TG}$ = 80 x 40 $nm^2$ and $W_{CH}$ x $L_{TG}$ = 1 x 1 $\mu m^2$ using the methodology described in the subsection B. Up to 6 devices per composition were tested, limited by the sample size fitting in the annealing tool chamber.

D. PBTI tests

PBTI tests were performed at 25 and 125°C using the eMSM scheme[39]. The higher temperature was chosen to enhance the H-driven $V_{TH}$ shift, which usually dominates at elevated temperatures[23]. The devices were stressed for $10^3$ s at overdrive voltages of $V_{OVD}$ = 1.2 - 2.4 V, and in one case for $10^5$ s (~40 hours) at $V_{OVD}$ = 2 V ($V_D$ = 50 mV in all cases), with a fresh device being used for each $V_{OVD}$. The overdrive voltage was defined as the difference between the gate bias stress ($V_{G_stress}$) and the device time-zero threshold voltage ($V_{TH0}$). The $V_{TH}$ shift was calculated by assuming a horizontal rigid shift of the $I_D$-$V_G$ curve after stress at fixed relaxation time of 1 s. $W_{CH}$ x $L_{TG}$ = 1 x 1 $\mu m^2$ devices were used in the tests to assess averaged trapping behavior.

E. *Ab initio* computations

All *ab initio* computations reported in this paper are performed using the CP2K software package version 8.2[40]. The hybrid Gaussian and plane wave density functional scheme of CP2K[41,42,43,44,45] makes that the dimension of the systems needed to reach low concentrations of defects are computationally feasible. We used the PBEsol generalized gradient approximation for the exchange correlation functional[46,47]. The standard double ζ valence plus polarization (DZVP) basis sets[48] and pseudo potentials[49,50,51] provided with CP2K are used. All calculations are performed using a single **k**-point (Γ), to prevent effects caused by the artificial periodicity introduced by the supercell approach to enter the results. For the structure optimization, we use a maximum geometry change convergence criterion of 5 mBohr and a force convergence criterion of 1 $mE_H$/Bohr. We use a target accuracy for the electronic self-consistency convergence of $10^{-6}$ $E_H$. The preparation, execution, monitoring, and post-processing of the over 2000 computations reported in this work have been facilitated by our in-house python package.

The amorphous structural models used in this work are generated using the decorate and relax method proposed by Drabold et al.[52]. In our experience, this approach leads to less defected structures at lower computational costs than melt and quench methodologies[53,54]. In each case we generate 10 super cell models of close to 420 atoms keeping the targeted stoichiometry. The structural optimization uses a combination of the Broyden–Fletcher–Goldfarb–Shanno (BFGS) algorithm[55,56,57,58] and time-stamped force-bias MonteCarlo (TFMC)[59,60].

The hydrogen binding energies are computed with respect to a neutral gas phase molecule, i.e. $H_2$. We first sample the original unit cell for sites where an interstitial hydrogen atom can potentially bind. The sampling is based on a 3D grid with a spacing of 1 Å in the unit cell, resulting in about 6000 points. A point on this grid is used if it is not closer than 1.1 Å to an existing atom. With this procedure close to 1000 sites are selected per system. For each of these sites the structure is fully optimized with that one additional hydrogen atom inserted.

For the hydrogen binding computations, we perform a direct 'local' minimization of the structure using the BFGS algorithm. This local optimization provides a clear picture of the energy distribution of the binding sites. We intentionally do not perform an *ab initio* TFMC or molecular dynamics simulated annealing type of optimization. This would collapse some of the more meta-stable binding sites into more stable ones. By performing a full, exhaustive, screening we will encounter the real global minimum by construction.

## II. RESULTS AND DISCUSSION

### A. Performance

Functional scaled devices were obtained with all tested In-poor IGZO compositions, see Fig. 2. All In-poor transistors showed a positive $V_{TH}$ directly after the fabrication without the need for $O_2$ anneal, while the In-rich sample required a 1 h, 250ºC $O_2$ anneal to bring $V_{TH}$ into the measurement range. The eliminated need for the $O_2$ anneal in In-poor devices allows to simplify the device layout in the future, by omitting the OT. This omission will enhance the device scalability and improve its compatibility with 3D integration. It should be noted that in this work all experiments were conducted on devices with OT.

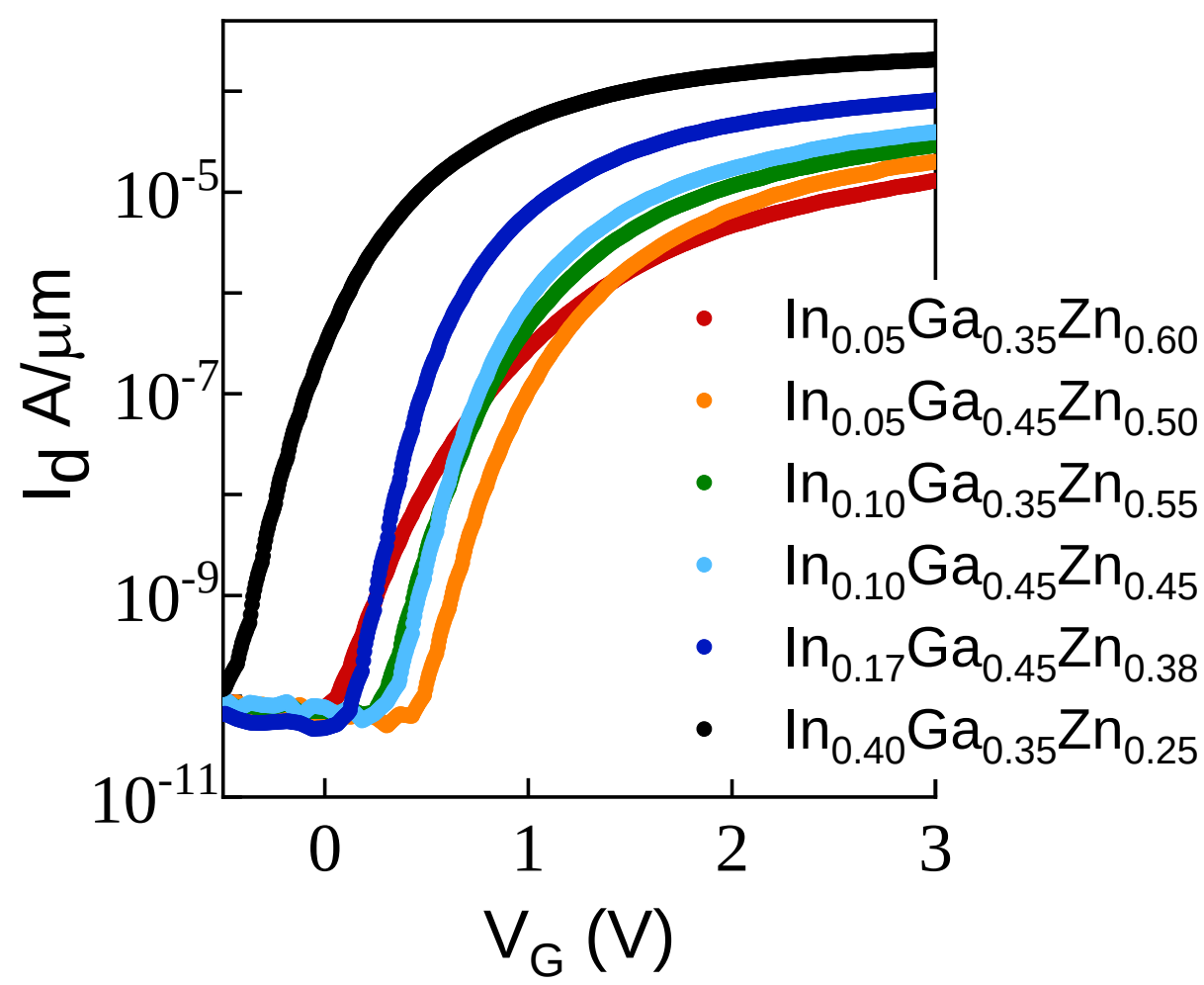

Fig. 2: The averaged transfer curves for TG FETs with different IGZO channel compositions (20 devices per composition). $W_{CH}$ x $L_{TG}$ = 80 x 40 $nm^2$, $V_D$ = 1 V.

A clear performance trade-off is observed for different IGZO compositions, where with increasing In and decreasing Zn concentrations µ, $R_C$, $I_{ON}$, and SS improve, but $V_{TH}$ degrades, Fig.3. The increase in In% is known to make IGZO less stable, leading to the decrease in the number of oxygen vacancies and thus in the carrier concentration[2, 61]. That in turn leads to higher film conductivity, improving µ, $R_C$, and $I_{ON}$, but degrading the $V_{TH}$. Such dependence on the IGZO composition is typical for IGZO and commonly reported in the literature[25, 61, 62]. However, the SS improves for the compositions for which $V_{TH}$ is degraded, which remains unclear, as the electrostatic control of the channel should degrade for higher carrier concentrations.

The best performance trade-off among In-poor devices is obtained with $In_{0.17}Ga_{0.45}Zn_{0.38}$, featuring median values of µ = 8 $cm^2/(V*s)$, $V_{TH}$ = 0.3 V, $I_{ON}$ = 17 µA/µm, SS = 94 mV/dec, $R_C$ = 19 kOhm*µm. Although the In-rich $In_{0.40}Ga_{0.35}Zn_{0.25}$ samples outperform the $In_{0.17}Ga_{0.45}Zn_{0.38}$ in terms of current conduction, the latter still has reasonable levels of mobility and $I_{ON}$, combined with as-fabricated positive $V_{TH}$.

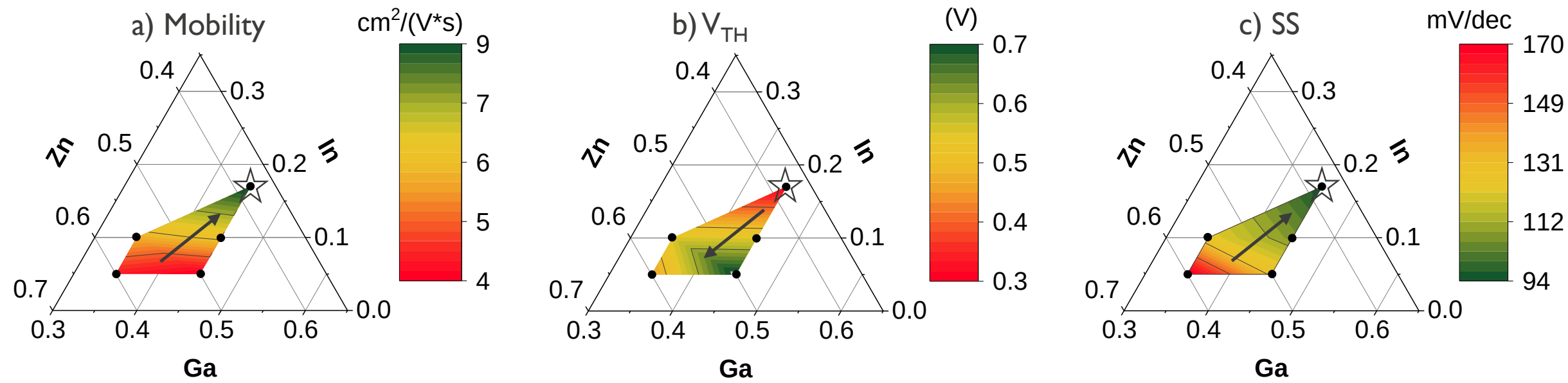

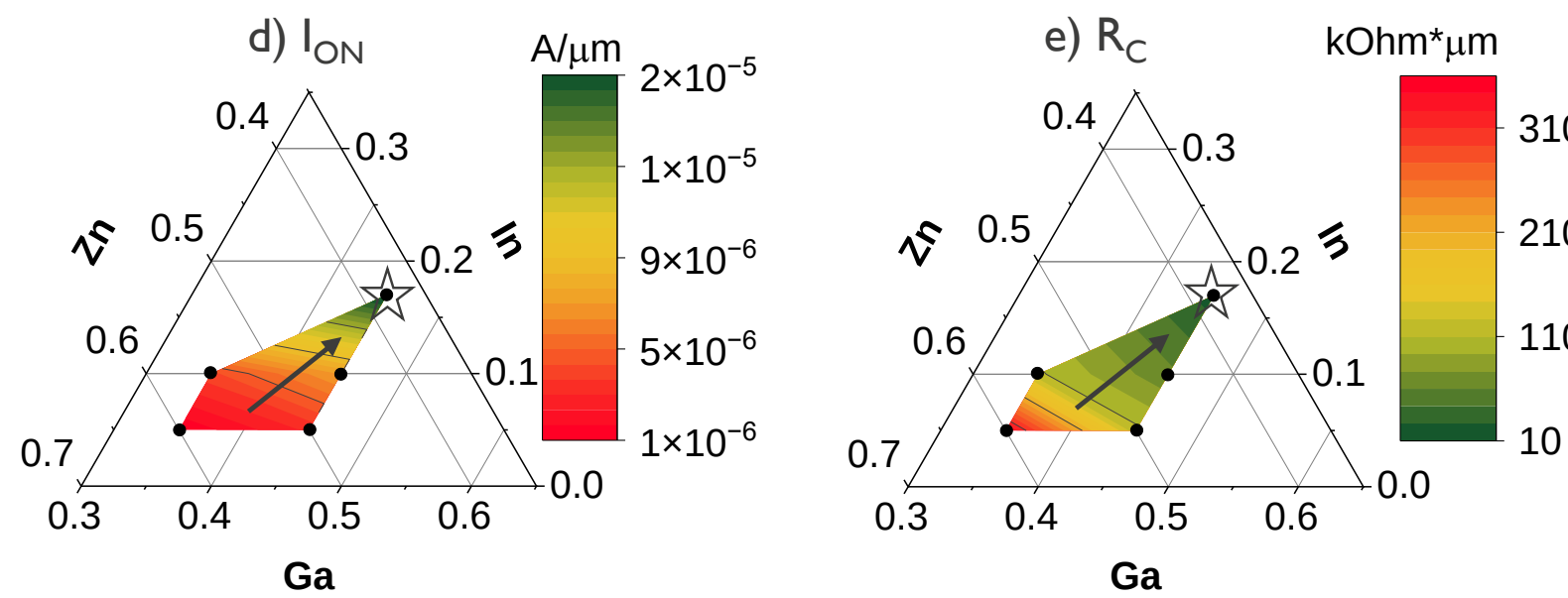

| f) | $In_{0.40}Ga_{0.35}Zn_{0.25}$ | $In_{0.17}Ga_{0.45}Zn_{0.38}$ |
|---|---|---|
| µ $cm^2/(V*s)$ | 19 | 8 |
| $V_{TH}$ (V) | -0.1 | 0.3 |
| SS (mV/dec) | 74 | 94 |
| $I_{ON}$ (µA/µm) | 28 | 17 |
| $R_C$(kOhm* µm) | 12 | 19 |

Fig. 3: Ternary diagrams depicting the composition dependence of µ (a), $V_{TH}$ (b), SS (c), $I_{ON}$ (d), and $R_C$ (e) for five In-poor IGZO channel transistors. A clear trade-off among different performance parameters is observed. Performance benchmarking of the In-rich and the best In-poor sample (f). $W_{CH}$ x $L_{TG}$ = 80 x 40 $nm^2$.

B. Resilience to H in PBTI

Several aspects of IGZO PBTI need to be highlighted before analyzing the experimental data. It is known that $V_{TH}$ shifts induced in IGZO devices during PBTI tests can be caused by 1) electron trapping in the gate dielectric causing positive $V_{TH}$ shifts; and 2) by H release from the gate dielectric into the IGZO channel resulting in negative $V_{TH}$ shifts[23]. These two PBTI mechanisms have very different dependencies on the stress voltage, stress time, and temperature[23,63]. The electron trapping process was reported to have a very weak dependency on time (time exponent $n \sim 0.1$), voltage (voltage acceleration factor $\gamma \sim 0.9$), and temperature (activation energy $E_a \sim 30$ meV). In contrast, the H-related degradation was shown to have a much stronger time ($n \sim 0.5$), voltage ($\gamma \sim 2.6$) and temperature ($E_a \sim 500$ meV) dependences[23]. With such a large apparent activation energy, the H de-trapping process appears to be reaction limited; in particular, we speculate that H is released by charging oxide traps[23,64]. In other words, a temperature stress ($T \leq 200$ºC) without the gate bias applied is not sufficient to release H species from the gate-dielectric and dope IGZO. As such a process is mediated by electron-injection, the strong voltage dependency becomes natural: the higher applied stress bias, the easier the injection, explaining the strong stress bias dependence of the H-induced $V_{TH}$ shift. Consequently, PBTI tests at high temperature (125ºC), high stress bias and prolonged time allow to detect the H component, if that is present in the device.

The device resilience to H in PBTI showed a strong dependence on the IGZO composition, see Fig. 4. The observed effects can be divided in three groups. First, the $V_{TH}$ shifts in $In_{0.05}Ga_{0.35}Zn_{0.6}$ are negative in the whole measurement range, Fig. 4a, revealing a very high sensitivity of this composition to H. Second, $In_{0.10}Ga_{0.35}Zn_{0.55}$, $In_{0.17}Ga_{0.45}Zn_{0.38}$, and $In_{0.40}Ga_{0.35}Zn_{0.25}$ initially show positive $V_{TH}$ shift; however, as the stress time increases, the trend reverses and $V_{TH}$ shift becomes negative, Fig. 4c-f. The $V_{TH}$ shift sign reversal indicates that both electron trapping (causing the positive $V_{TH}$ shift) and H release from the gate oxide into the channel (responsible for the negative $V_{TH}$ shift) are present in the devices and dominate at different stress time and voltage. Third, $In_{0.05}Ga_{0.45}Zn_{0.50}$ shows only positive $V_{TH}$ shifts, even after 40 h of stress, Fig. 4b,g, demonstrating the resilience of this composition to H.

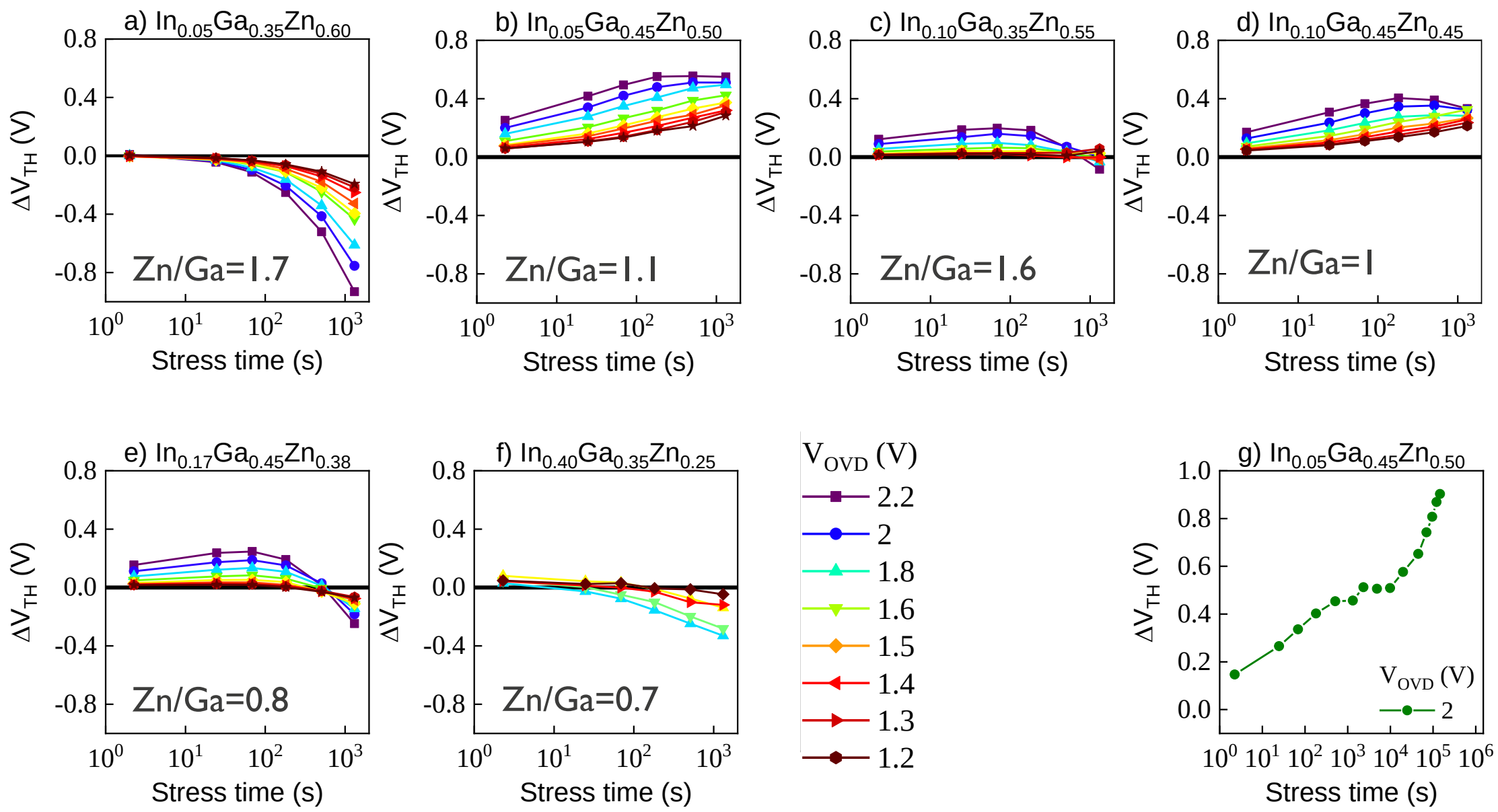

Fig. 4: PBTI test at 125ºC results for devices comprising channels with different IGZO compositions (a-g). The H-induced shift component is reduced in the best performance In-poor sample $In_{0.17}Ga_{0.45}Zn_{0.38}$ (e) compared to In-rich $In_{0.40}Ga_{0.35}Zn_{0.25}$ (f, h), and is undetectable within the measurement window in $In_{0.05}Ga_{0.45}Zn_{0.50}$ (b) even after 105 s (~ 40 hours) of stress (g). $W_{CH}$ x $L_{TG}$ = 1 x 1 $\mu m^2$, $V_D$ = 0.05 V.

Additional experimental evidence that supports the absence of the H-doping process in the $In_{0.05}Ga_{0.45}Zn_{0.50}$ devices is the fact that for this composition the positive $V_{TH}$ shift at 125ºC is always larger than at the positive $V_{TH}$ shift at 25ºC, as shown in Fig. 5 (a). The Non-Radiative Multiphoton model (NMP) predicts that electron trapping is enhanced at higher temperatures[65]. Therefore, if the $\frac{\Delta V_{TH_125ºC}}{\Delta V_{TH_25ºC}}$ ratio is larger than 1 at the same stress condition, then the H-doping process is not measurable within the experimental window. A ratio with a value between 0 and 1 indicates an apparent negative activation energy, which means that the degradation at 125ºC, although still positive, is smaller in magnitude than at 25ºC. Such behavior can only be explained by an electron trapping-dominated process, but with a non-negligible presence of H-doping mechanism at 125ºC. Finally, a negative ratio indicates that the H-doping process dominates at 125ºC, inducing a measurable negative $\Delta V_{TH}$. As can be seen from the Fig. 5, pure electron trapping occurs only in $In_{0.05}Ga_{0.45}Zn_{0.50}$. $In_{0.10}Ga_{0.45}Zn_{0.45}$ reveals a non-negligible hydrogen component at high overdrives, while for other compositions the H-doping process dominates the degradation inducing a measured negative $\Delta V_{TH}$ at 125ºC, especially for $In_{0.05}Ga_{0.35}Zn_{0.60}$ also $In_{0.40}Ga_{0.35}Zn_{0.25}$. Consequently, the In-poor compositions (except for $In_{0.05}Ga_{0.35}Zn_{0.60}$) demonstrate higher resilience to H than the In-rich IGZO. However, a reliability-performance trade-off is present, as $In_{0.05}Ga_{0.45}Zn_{0.50}$ that showed the highest resilience to H offers the lowest performance (Fig. 3), which might however be improved by further contacts optimization and the channel thickness increase.

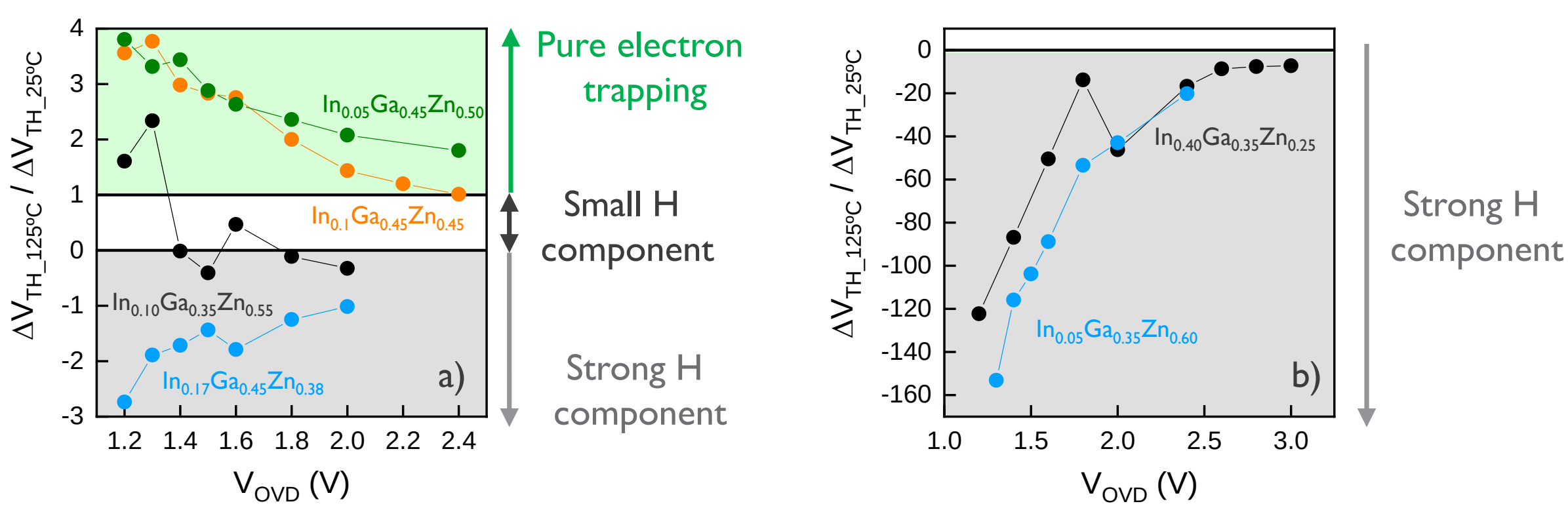

Fig. 5: The ratios of $V_{TH}$ shifts measured in PBTI tests at 125ºC and 25ºC as a function of overdrive voltage and IGZO composition. For better visualization the data are split in two graphs, with (a) capturing the compositions with weaker H doping and (b) - with stronger H doping.

The experimental data indicates that lower In content and lower Zn/Ga ratio increase the device resilience to H in PBTI, which follows from the following observations. $In_{0.05}Ga_{0.35}Zn_{0.60}$ and $In_{0.05}Ga_{0.45}Zn_{0.50}$ showed high sensitivity and high resilience to H, respectively. As they have the same In content, lower Zn/Ga ratio must lead to higher resilience to H in PBTI. However, the decrease of the Zn/Ga ratio alone is not sufficient for achieving the resilience to hydrogen in PBTI, as $In_{0.17}Ga_{0.45}Zn_{0.38}$ and $In_{0.40}Ga_{0.35}Zn_{0.25}$ have even smaller Zn/Ga ratios (0.7 and 0.8) than $In_{0.05}Ga_{0.45}Zn_{0.50}$ (1.1), but they clearly show the signs of the H-induced PBTI component, the $\Delta V_{TH}$-stress time trend reversal and the negative net $V_{TH}$ shift values.

The composition-dependent sensitivity to H in PBTI tests can be explained through different stability of $InO_x$, $GaO_x$, and $ZnO_x$. As stated previously, in the H-induced $V_{TH}$ shift in PBTI, the H gets released from the gate-dielectric, migrates to the channel and finally incorporates into it by ionically bonding to the oxygen, taking one of the four coordination sites at an oxygen atom at the cost of an In, Ga, or Zn ion. As InOx and ZnOx are less stable than GaOx[61,66,67], H binding is less stable in In- or Zn-poor and Ga-rich IGZO, decreasing the H-driven PBTI component in such compositions.

C. Resilience to hydrogen in FGA

In-poor IGZO shows superior resilience to H in harsh annealing conditions, as compared to the In-rich IGZO, see Fig. 6. All In-poor compositions survive 1 + 1 h FGA at 420ºC with very small changes in performance, while In-rich IGZO experiences strong negative $V_{TH}$ shifts and the increase in drain current. Such a change in performance is a typical manifestation of the n-type H-induced IGZO channel doping[18]. Remarkably, In-poor IGZO samples showed only small difference in stability to H in FGA, while their stability to H in PBTI varied significantly depending on the composition, Fig. 4. No significant nor systematic impact of $V_D$ and device dimensions on the devices stability in FGA was observed, Fig. 7.

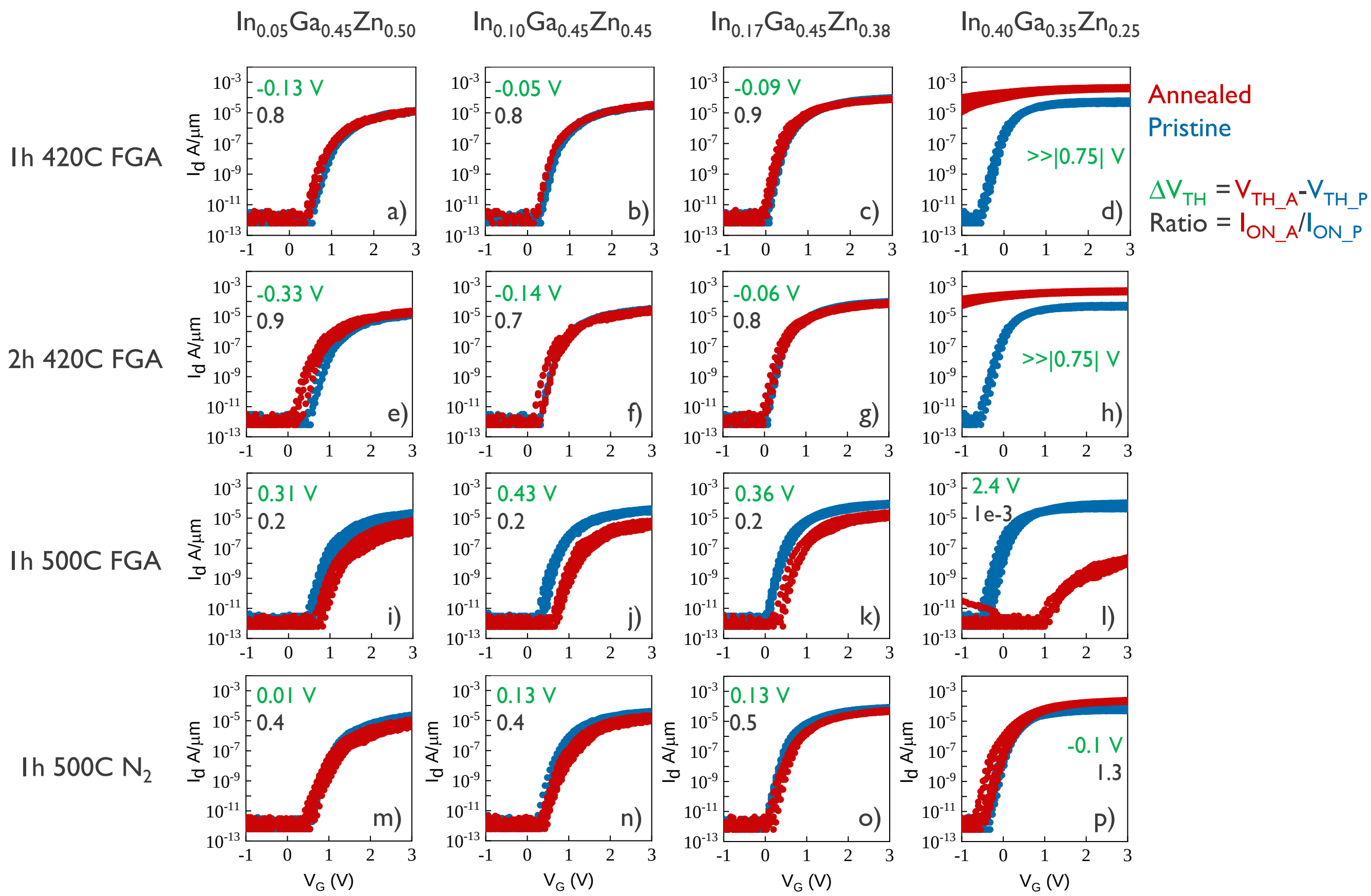

Fig. 6: Impact of different FGAs ($H_2 = 10\%$) (a-l) and $N_2$ anneal (m-p) on performance of In-poor and In-rich IGZO channel devices. In-poor channels demonstrate much higher stability in FGA than the In-rich one. Higher stability of all samples in $N_2$ anneal compared to that in FGA at the same temperature indicates that H, and not only the temperature, impacts the device stability. The numbers indicate $V_{TH}$ shift relative to the initial $V_{TH}$ (top number) and the ratio of the on-current of the annealed sample to that of the pristine one (bottom number). All samples were annealed simultaneously to exclude the impact of the annealing process variability. $W_{CH} \times L_{TG} = 80 \times 40\ nm^2$, $V_D = 1$ V.

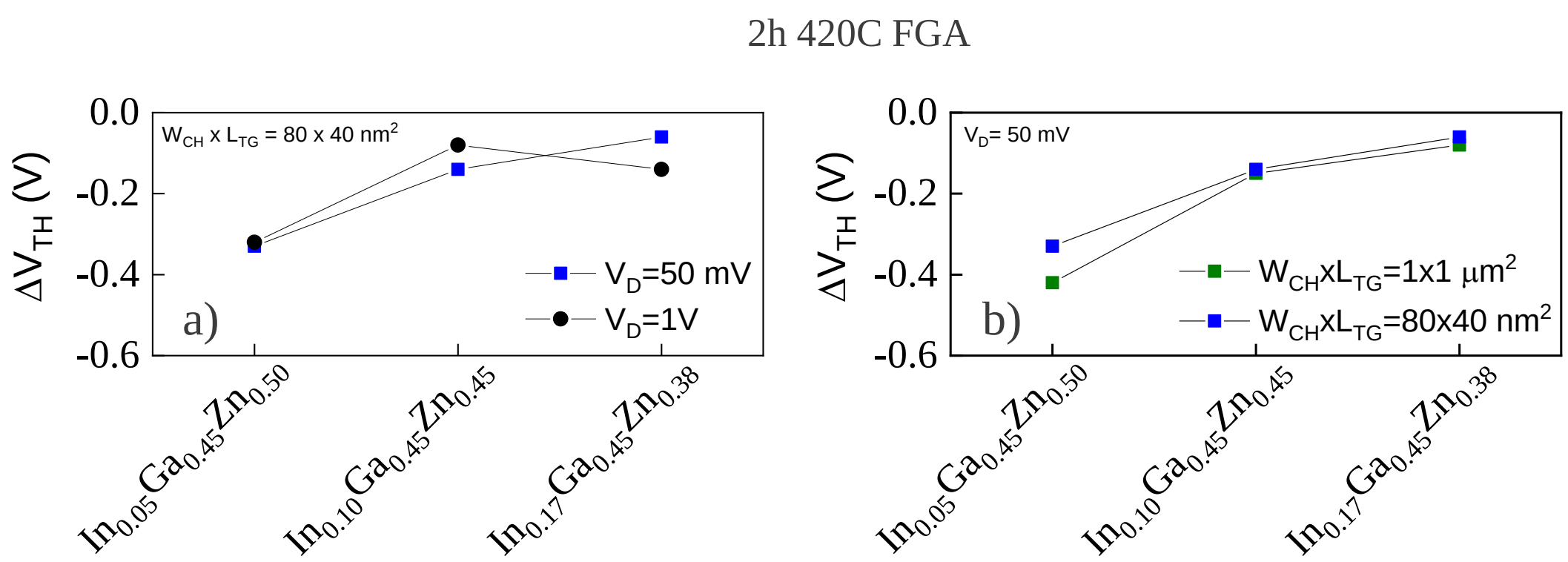

Fig. 7: Median $V_{TH}$ shifts after 2 h 420ºC FGA (calculated by subtracting the $V_{TH}$ of the pristine devices from the $V_{TH}$ of the annealed samples). No significant nor systematic impact of a) $V_D$ voltage and b) device dimension on the device stability in FGA is observed. Small differences in values are attributed to natural variability present in the samples.

After the 1 h at 500ºC anneal, all In-poor and In-rich samples show positive $V_{TH}$ shifts and decrease in the drain current, see Fig. 6i-l. While this change is rather mild in the case of the In-poor devices, whose functionality is preserved, In-rich samples are strongly degraded. At the same time, all samples (including the In-rich one) showed much smaller change in the performance after 1h 500ºC annealing in $N_2$, compared to FGA at the same temperature, Fig. 6i-p. This indicates that the presence of H, and not only the elevated temperature, has a strong impact on the device stability. The 500ºC FGA disrupts the channel integrity, as could be seen from TEM images for the In-rich $In_{0.40}Ga_{0.35}Zn_{0.25}$ and one of the In-poor $In_{0.17}Ga_{0.45}Zn_{0.38}$ examples, see Fig. 8. While Ga atoms remained relatively stable, forming a continuous layer (Fig. 8b, f, i, j), Zn and especially In atoms diffused, leading to the formation of the material clusters and voids (Fig. 8c, d, g, h, i, j). This agrees well with the fact that higher Ga% improves IGZO stability, while higher In and Zn decrease it[61]. The migration was more pronounced in the In-rich sample, resulting in a highly resistive channel predominantly composed of Ga and some Zn, which agrees with the severe drain current degradation and positive $V_{TH}$ shift, see Fig. 6l. Although migration was also observed in the In-poor sample, a greater amount of Zn and In remained within the channel, leading to formation of regions of higher conductance interspersed with more resistive Ga-rich areas, which aligns with the moderate drain current degradation and $V_{TH}$ change, see Fig. 6k. The provided explanation is based on the knowledge that higher Ga% decreases IGZO conductivity and leads to more positive $V_{TH}$, while higher Zn and especially In content have the opposite effect[61].

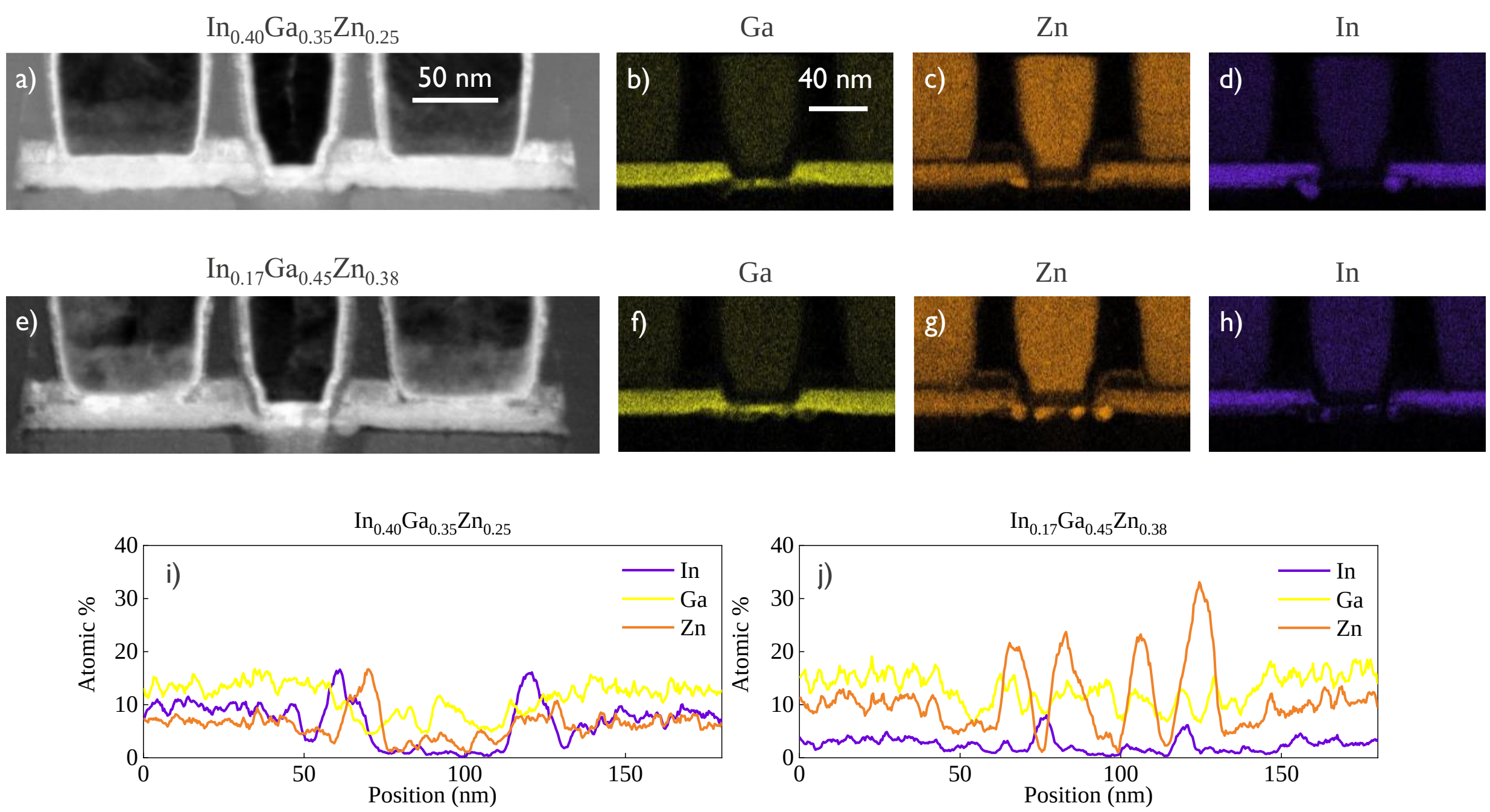

Fig. 8: TEM images, composition maps, and line scans of $In_{0.40}Ga_{0.35}Zn_{0.25}$ (a-d, i) and $In_{0.17}Ga_{0.45}Zn_{0.38}$ (e-h, j) samples that were annealed for 1h 500ºC FGA. $W_{CH}$ x $L_{TG}$ = 80 x 40 nm$^2$.

By comparing the impact of FGA at 420ºC and 500ºC, it becomes evident that there is a certain temperature threshold, below which H dopes IGZO and thus affects its electrical performance; above this threshold the interaction leads to channel disintegration. The following section will focus on the first regime and investigate the mechanisms guiding IGZO resilience to H in PBTI and FGA tests.

D. Mechanisms of the H impact in FGA

Using *ab initio* and thermodynamic calculations, we identify two mechanisms of H interaction with IGZO: 1) O scavenging[68] and 2) H incorporation[52]. In the first mechanism (Fig. 9a), O scavenging reaction produces $H_2O$ and donates two electrons to IGZO. How easily O can be scavenged depends on the stability of the IGZO composition, which increases with lower In%[61]. In the second mechanism, incorporated H can reside near O (Fig. 9b) or near metals (In, Ga, Zn, Fig. 9c), donating or accepting electrons, respectively. *Ab initio* computations show (Fig. 10) that for both In-rich and In-poor IGZO, it is energetically more favorable for H to reside near O than near a metal, resulting in a net electron donation.

Thermodynamic considerations show that increasing temperature promotes O scavenging and demotes H incorporation, Fig. 11. This happens because the O scavenging reaction increases the system entropy, as the produced in the reaction $H_2O$ (green line in Fig. 11) has higher entropy than the initially present $H_2$ (grey line in Fig.11). Moreover, the entropy gain obtained in the scavenging reaction increases with the higher temperature (blue line in Fig. 11). Opposite to that, H incorporation in IGZO leads to entropy decrease, as the gas species embeds in the solid and no other gas is produced. The entropy loss becomes higher with the increasing temperature (grey line in Fig. 11), leading to H incorporation being demoted by higher temperature.

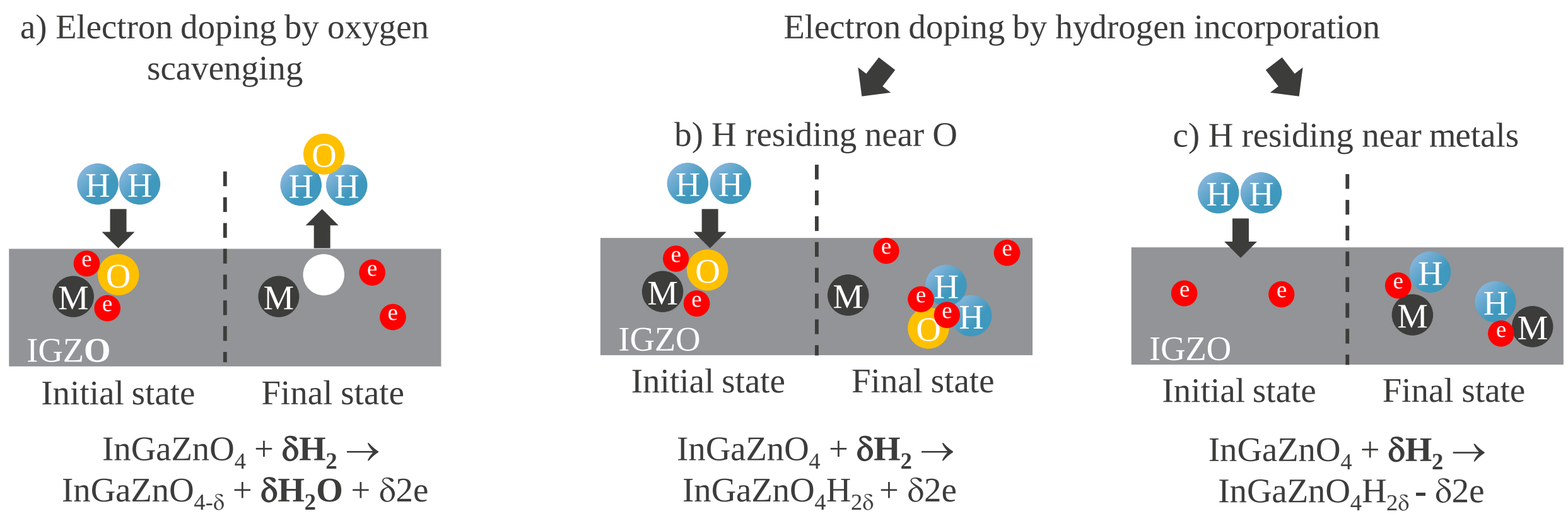

Fig. 9: Three mechanisms of IGZO doping by H: a) electrons donation via O scavenging, b) electrons donation by H incorporation and residing near O, and c) electrons acceptance via H incorporation and residing near metal (In, Ga, Zn). In the case of H incorporation, H atom, which initially has one electron in its orbital can either accept or donate one electron, because either process allows to complete the orbital and thus minimize the energy.

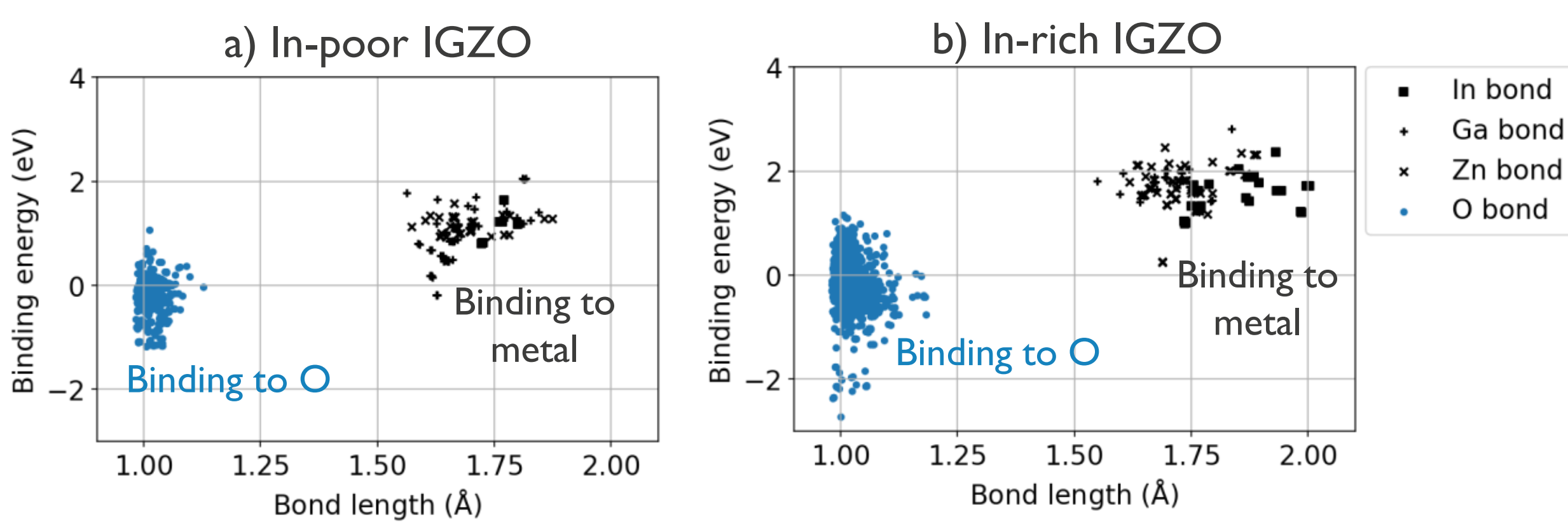

Fig. 10: *Ab initio* computed binding energy of H when residing near metal (In, Ga, Zn) or O ions, for In-rich (a) and In-poor (b) IGZO. In both types of IGZO, neighbouring O is more energetically favourable than neighbouring metal ions, leading to H incorporation having the net electrons donation effect on IGZO.

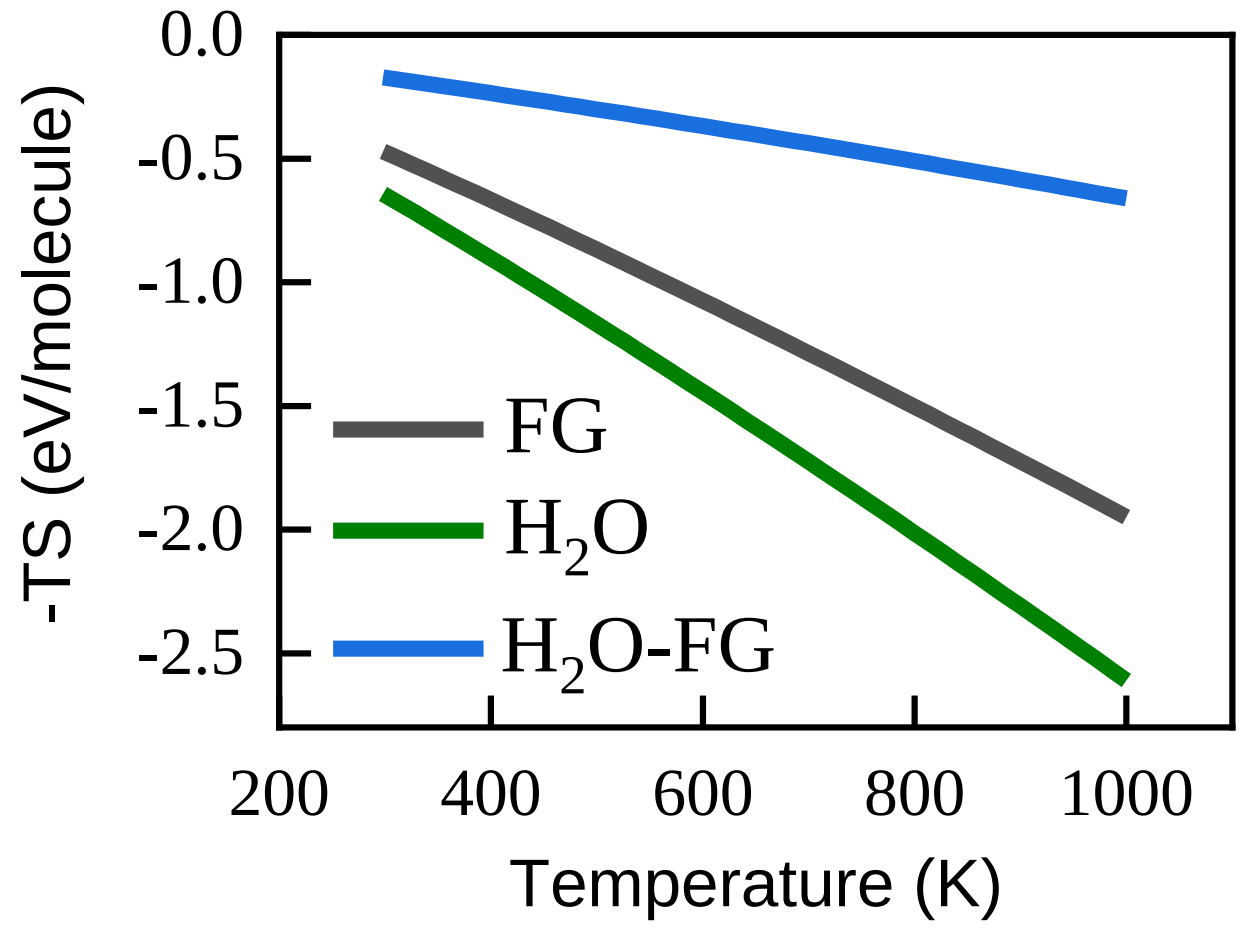

Fig. 11: Gas phase entropy as a function of temperature for forming gas (black) and $H_2O$ (green). With the increase in temperature, O scavenging by H is promoted (see also Fig. 10), as conversion of $H_2$ into $H_2O$ leads to the increase in entropy making Gibbs' free energy more negative.

Based on the above considerations, we argue that the following mechanism guides resilience to H in FGA. FGA happens at higher temperature (420ºC), and H is present in the ambient in the gas phase. As the higher temperature demotes H incorporation and promotes O scavenging, we argue that at FGA temperatures O scavenging is more dominant than H incorporation. Since In-poor films are more stable than In-rich IGZO films[61], they show higher resilience to O scavenging and thus are more stable in FGA. It should also be noted that different mechanisms guide the IGZO-H interaction in FGA and PBTI tests (which also happen at different temperature and bias conditions), the resilience to H in FGA does not automatically translate into the resilience to H in PBTI.

## IV. CONCLUSION

We demonstrated that the use of In-poor IGZO as a channel material allows to achieve superior stability to H in PBTI and FGA, compared to the In-rich IGZO. After a 2 h, 420ºC FGA, scaled In-poor top gate devices showed only a small change in performance, while the In-rich channel devices lost modulation in the measurement window. Most of the In-poor devices demonstrated and improved resilience to H in PBTI compared to In-rich samples, and $In_{0.05}Ga_{0.45}Zn_{0.50}$ demonstrated no sign of the H-induced $V_{TH}$ shift even after 40h of stress time. An IGZO composition-dependent trade-off between the performance and resilience to H in FGA and PBTI was observed, with $In_{0.17}Ga_{0.45}Zn_{0.38}$ offering a good compromise. We also showed that IGZO interaction with H in FGA and PBTI is guided by different processes, being oxygen scavenging by H and H release from a gate-dielectric into the channel, respectively, and resilience to H in one does not necessarily translate to resilience to H in the other one. The demonstrated high resilience of In-poor IGZO to H makes IGZO transistors compatible with H treatments and 3D integration, improving their potential for industrial production.

## ASSOCIATED CONTENT Supporting Information

Device fabrication details

## ACKNOWLEDGEMENT

This work was funded by the imec industry-affiliated Active Memory program and has been enabled in part by the NanoIC pilot line. The acquisition and operation are jointly funded by the Chips Joint Undertaking, through the European Union's Digital Europe (101183266) and Horizon Europe programs (101183277), as well as by the participating states Belgium (Flanders), France, Germany, Finland, Ireland and Romania. For more information, visit nanoic-project.eu. We thank imec FAB, Amsimec, and MSP teams.

Graphic for manuscript

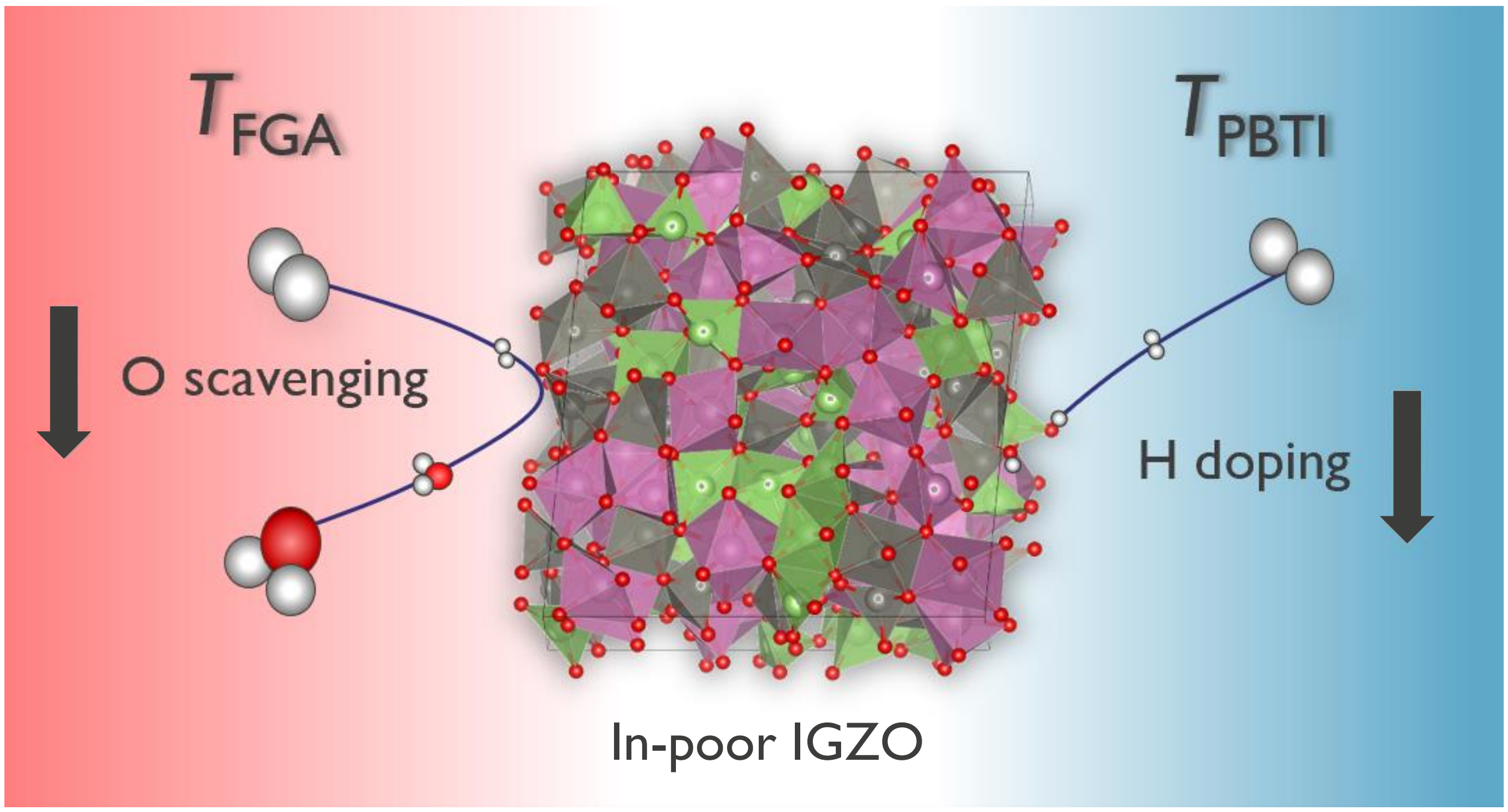